# An evolutionary view on the emergence of Artificial Intelligence

Matheus Eduardo Leusin[1], Björn Jindra[1,2], and Daniel S. Hain[3]

[1] Faculty of Business Studies and Economics, Chair of Innovation and Structural Change, University of Bremen, Germany.
[2] Department of International Economics, Government and Business, Copenhagen Business School, Denmark
[3] AI: growth Lab, Aalborg University Business School, Denmark

**Abstract:** This paper draws upon the evolutionary concepts of technological relatedness and knowledge complexity to enhance our understanding of the long-term evolution of Artificial Intelligence (AI). We reveal corresponding patterns in the emergence of AI - globally and in the context of specific geographies of the US, Japan, South Korea, and China. We argue that AI emergence is associated with increasing related variety due to knowledge commonalities as well as increasing complexity. We use patent-based indicators for the period between 1974-2018 to analyse the evolution of AI's global technological space, to identify its technological core as well as changes to its overall relatedness and knowledge complexity. At the national level, we also measure countries' overall specialisations against AI-specific ones. At the global level, we find increasing overall relatedness and complexity of AI. However, for the technological core of AI, which has been stable over time, we find decreasing related variety and increasing complexity. This evidence points out that AI innovations related to core technologies are becoming increasingly distinct from each other. At the country level, we find that the US and Japan have been increasing the overall relatedness of their innovations. The opposite is the case for China and South Korea, which we associate with the fact that these countries are overall less technologically developed than the US and Japan. Finally, we observe a stable increasing overall complexity for all countries apart from China, which we explain by the focus of this country in technologies not strongly linked to AI.

# 1. Introduction

During the last decade, we have been witnessing an all-time high of publications and inventions related to Artificial Intelligence (AI) (WIPO, 2019) due to the development of high-performance parallel computing chips and large datasets that extended the applicability of AI technology (Klinger et al., 2018; Li & Jiang, 2017). From a historical perspective, this is not the first surge in AI development. Since its inception in the late 1950s, AI had at least two periods characterised by increased research and investment, which ultimately stalled (Li & Jiang, 2017; WIPO, 2019).

Today, there are high expectations as well as concerns about the development of AI and its economic impact. At the macro-level, countries mastering AI early are expected to gain advantages over global markets and industries (Cave & ÓhÉigeartaigh, 2018; Cockburn et al., 2018; Klinger et al., 2018). At the micro-level, firms failing to adopt AI are expected to lose significant market shares to early adopters or new entrants, with some incumbents becoming obsolete (European Commission, 2017). AI might also change the nature of scientific and technical advancement itself (Cockburn et al., 2018). Accordingly, countries are boosting their investments in AI development, rushing into a so-called 'AI global race' (Klinger et al., 2018).

Prior research reveals distinct patterns of how countries develop AI. In principal, the US and China appear to lead in terms of AI outputs. For the US, a market-oriented AI development is emphasized, whereas in the case of China the government plays a more important role and universities are the main AI developers (Fujii & Managi, 2018; Righi et al., 2020). In terms of patterns of intellectual property protection, it seems that the US focuses upon the exploration of AI in international markets, whereas China prioritizes the domestic market (Leusin et al., 2020). Finally, the dominance of AI by leading countries such as the US and China might close the window of opportunity in AI to latecomers, due to long-lasting early adopter advantages (Klinger et al., 2018).

In this paper, we investigate the emergence of AI by focusing on the bidirectional evolutionary link between technology dynamics and countries' technological capabilities as theoretical contribution. In particular, we use the concepts of technological relatedness and knowledge complexity to analyse the patterns of the long-term development of AI as well as the potential impacts of this development at the country-level.

Firstly, we propose that AI emergence at the global level is associated with increasing related variety due to knowledge-relatedness as well as increasing complexity. While related variety potentially converts AI into a specialised technology, which demands high efforts from firms and other actors developing it, complexity relates to potentially higher economic benefits resulting from its exploration. Due to path dependency, we expect that the emergence of AI is also influenced by countries' technological capabilities, which create distinct patterns of technological development. In turn, AI emergence is also likely to affect countries' technological development proportionally to the progress in AI. As AI evolves, the effect of developing AI becomes stronger: countries developing it will also be creating increasingly specialised and profitable technologies. Thus, secondly, we propose that AI's evolution affects the technological paths of countries at the technological frontier of AI by increasing the overall relatedness and complexity of innovations developed by these countries.

To explore these two propositions, we analyse priority patents registered in the observation period 1974-2018. We use inventors' location information as a proxy for identifying the countries related to each patent (De Rassenfosse et al., 2019a), together with a keyword-based approach to identify AI patents registered within this period. We analyse AI emergence at the global and country-level. At the global level, we use existing technological classifications to identify the technological core of AI, which we differentiate from related and surrounding technological fields during the observation period. At the country level, we analyse how the technological specialisations of countries at the AI frontier (US, Japan, South Korea, China) possibly affect the geographic emergence of AI. We use a technological space approach, adapted from Hidalgo et al. (2007), to analyse the process of AI emergence and calculate indicators of relatedness and knowledge complexity, used to measure both the development of AI and the technological development of countries.

The following section provides the theoretical background related to the emergence and evolution of technologies and discusses the relatedness and complexity concepts. The data collection and methodology are detailed in Section 3. Section 4 presents the empirical analysis. In the first part of this section, we investigate the emergence of AI at the global level; in the second part, we analyse the impacts of AI emergence at the country level. Finally, section 5 discusses the main findings, implications, and limitations.

## 2. Theoretical background

*2.1. AI from a general purpose technology perspective*

The issues on the emergence and effects of technological development are addressed in the literature from a variety of theoretical perspectives. In the case of AI, much of the debate focuses on the question of whether it is a General Purpose Technology (GPT), or not (Brynjolfsson et al., 2018; Cockburn et al., 2018; Klinger et al., 2018; Lipsey et al., 2005; Trajtenberg, 2018). The notion of GPT refers to technologies that are pervasive, present an inherent potential for technical improvements over the years, and induce the creation of related 'innovational complementarities' (Bresnahan & Trajtenberg, 1995).

In principal, this theory focuses on the economic impacts of GPTs in terms of adoption and diffusion processes across firms and sectors, rather than the evaluation of the technological development itself. This perspective requires a measure of the technology's pervasiveness. Yet, due to the modularity of AI, which potentially allows embedding it in any other technology able to function with the foresight of its environment (Nilsson, 2009), its pervasiveness is particularly hard to measure. AI is also understood as a transversal technology meaning that it cannot be identified as a part of a specific sector, but rather it is a composition of technological fields without precise boundaries (Righi et al., 2020). It is also important to highlight that traditional GPT theory fails to shed light on the emergence of such technologies, mainly because of the difficulty in modelling GPTs by another way than arriving from outside of the considered system (Cantner & Vannuccini, 2012). Therefore, the GPT framework is not adequate for capturing the precise nature of the AI perspective (Vannuccini & Prytkova, 2020).

*2.2. AI from an evolutionary perspective*

2.2.1. The geographical development of technologies

Alternatively, the evolutionary perspective considers how distinct geographies affect technological development through the micro and meso dynamics that affect firms' behaviour. This view stresses that differences between geographical spaces, such as countries, produce different types of innovation. The main reason for that is the cumulative, path-dependent, and interactive process which characterises the creation of knowledge (Dosi, 1982; Nelson & Winter, 1982). This process produces geographical agglomerations of particular technologies, capital, institutions, and skills, which in turn, shape the creation of innovations differently (Freeman, 1987; Lundvall, 1992). Concurrently, these innovations also affect countries' technological development, since the emergence of technologies provides a new context for extant technologies to be recombined, generating new variations of innovation (Castaldi et al., 2015).

From this perspective, the local development of AI is both critical for creating related innovations and an option for sustaining further technological development of the exploring countries. Yet, despite the extensive literature addressing separately countries' innovation systems development and the emergence of new technologies (Nelson, 1992; Rakas & Hain, 2019; Uriona-Maldonado et al., 2012), the link between the technological dynamics of a long-time evolving technology and countries technological development remains underexplored.

2.2.2. Relatedness and complexity perspectives in technology evolution

The evolutionary perspective also considers in depth how new technologies emerge and evolve. For example, Breschi et al. (2003) suggest that knowledge-relatedness is a central factor affecting firms' technological diversification. The concept encompasses three dimensions of knowledge: i) knowledge proximity, related to learning processes from unintended learning spillovers or intended local learning; ii) knowledge commonalities, related to the possibility of firms' innovative activities spanning over more than one technology as a result of the same type of knowledge being used in more than one technology; and iii) knowledge complementarities, related to firms' need to master more than one technology for being able to develop new products and/or processes. In essence, firms extend their innovative activities across knowledge-related technological fields as a consequence of both unintended and intended learning processes and of knowledge features of relatedness and its links.

These micro-dynamics at the firm-level also generate macro-effects at the country level. Hidalgo et al. (2007), for example, emphasise that relatedness between products affects how countries explore new goods. The authors show that, over time, countries move through a 'product space' by developing goods close to those for which they already have established some comparative advantage, whereas more distant products are hardly reached. Furthermore, technologies used to produce goods also have different levels of complexity (Hidalgo & Hausmann, 2009). The most profitable ones demand more diverse and complex capabilities and are hence produced by a reduced number of countries, while the simplest goods demand few and less complex capabilities and technologies, and thus can be produced by more countries.

Other than products, the 'knowledge space' is used, for representing the relatedness between technologies (Balland, 2016). For instance, Petralia et al. (2017) highlight that countries are more likely to move into related technologies and that this behaviour is particularly strong in the earlier stages of countries' technological development. The authors also demonstrate that this effect diminishes over time and countries become less dependent on relatedness, moving into more complex technologies as they become more developed. At the technological level, there seem to be no significant differences in the likelihood of moving to more complex technologies.

Since knowledge dynamics are a higher-order reflection of micro-level dynamics and behaviour of firms and other organizations (Balland, 2016), the knowledge space framework is also commonly focused in smaller geographical spaces, such as regions and cities (Balland et al., 2019; Boschma et al., 2014; Rigby, 2015). In short, these investigations show that relatedness is a key measure for explaining the costs and likelihood of moving from one type of knowledge to another, while complexity provides a way of assessing the potential benefits of such movement. However, there is a missing link when explaining how countries are able to move successively to more complex technologies (Petralia et al., 2017), if at the micro-level firms explore new technologies uniquely according to knowledge-relatedness (Breschi et al., 2003).

Another important element for understanding technological development and the emergence of technologies comes from the 'recombinant innovation' process, which produces technological variety. Van den Bergh (2008) highlights that variety denotes the number of different technologies comprised of different functions in a population of elements. Frenken et al. (2007) emphasise the difference between variety as a source of regional knowledge spillovers and variety as a portfolio protecting a region from external shocks. The former is best measured by related variety (within sectors), while the latter is better captured by unrelated variety (between sectors).

Building upon this idea of recombinant innovation, Castaldi et al. (2015) and Solheim et al. (2018) find evidence that related variety enhances innovation by allowing technologies to be more easily recombined, whereas unrelated variety enhances technological breakthroughs by connecting previous unrelated technologies. Castaldi et al. (2015), in particular, stress that the emergence of new, recombinant technology provides a new context for extant technologies to be recombined, hence potentially allowing previously unrelated technologies to become related. More generally, Frenken et al. (2012) argue that the creation of variety, which can be fused through recombination, acts as a technological 'short-cut', supporting technological transitions that would be hardly accomplished otherwise.

2.2.3. Relatedness, complexity, and AI emergence

Against this background, we employ the relatedness and complexity perspectives as conceptual bases for understanding the emergence and development of AI. In particular, we argue that the development of a modular recombinant technology such as AI allows increasing the agglomeration of knowledge commonalities (i.e., the technological knowledge can be combined with even more technologies) and complementarities (i.e., actors will share a common kind of knowledge), resulting in an increase in the overall relatedness of a country's inventions. Similarly, the recombination of AI with other technologies will produce related variety, bridging a wider range of related technological opportunities for firms. This process is to be particularly enhanced in the case of AI due to its transversal characteristic (Righi et al., 2020), which allows applying it within a variety of technological sectors. The resulting technological bridges created by this process allow combining previously unrelated technologies, increasing the overall complexity of a country's inventions.

Furthermore, we analyse to which extent countries' process of developing an emerging modular technology like AI affects the relatedness and complexity of innovations associated with this technology. In particular, we expect that, due to path-dependency, countries will produce increasingly related innovations, thus increasing the overall relatedness of AI innovations. Moreover, we expect that the further technological development of AI will increase the overall complexity of AI innovations, ultimately making such innovations also more profitable.

## 3. Data and method

The history of AI's technological development recently culminated in the boost of AI-related innovations (Klinger et al., 2018; WIPO, 2019). Tracing back AI's development from the recent decade will allow us to analyse its emergence possible effects on countries' technological development. To measure that, we firstly apply the technological space framework to patent data investigating how AI evolved between 1974 and 2018, and secondly, we analyse how countries extensively developing AI technologies changed their technological spaces in the same observation period.

*3.1. Data collection and identification*

We use patent data from PATSTAT 2019 (Spring version). The generation of our patent dataset rests on two main choices: i.) the strategy for identifying AI patents, and ii.) assignment of patents to countries. For the former, we use a keyword-based search strategy, identifying as AI-patents all priority filings[1] (granted or not) mentioning at least one typical AI technique in their title or abstract. For selecting these AI-techniques we use the classification proposed by WIPO (2019), which is based on a computing classification scheme. We complement the 21 keywords for AI-techniques with their respective synonyms collected from Wikipedia. The resulting search strategy has a total of 36 keywords (see Appendix A).

---

[1] A priority filing is the first patent application filed to protect an invention. If the same patent is registered in other patent offices, the following registrations are called non-priorities, constituting a patent family linked through the priority filing.

In terms of geography, we follow De Rassenfosse et al. (2019a) by using the inventors' location as a proxy for identifying the countries to which each patent is related. An alternative choice would be to use assignees to determine a patent's country assignment. However, this would capture the origin of property rights rather than knowledge creation (Squicciarini et al., 2013). We adapt the strategy proposed by De Rassenfosse et al. (2019a) to identify source countries in the 2019 PATSTAT (Spring version), since the original dataset presented by the authors ends in 2014, which would potentially omit important AI developments that occurred thereafter. As De Rassenfosse et al. (2019a) use additional sources to increase the location information of inventors and applicants related to their dataset, we perform a cross-check validation with our dataset. We find that 23,983 of our 42,971 AI-priorities are registered between 2015 and 2018. Of the remaining 18,988 priorities, 15,355 are also found in the dataset presented by De Rassenfosse et al. (2019b). These patents hold the location information for 28,324 inventors, from which our dataset contains the same information for 27,275 inventors, meaning a 96.3% correspondence with the authors referred work.

In addition to this 'AI dataset', we also create a 'General dataset', composed of all 30,000,524 priority filings registered in the same period (1974- 2018), to which we apply the same strategy for identifying the source countries.

*3.2. Method*

After identifying AI priority fillings using a keyword-based search, we adopt the International Patent Classification (IPC) scheme for analysing the results. We calculate spatial indicators (e.g., location quotient) and co-occurrence (adjacency) matrices between entities, and indicators of relatedness and knowledge complexity[2]. Relatedness measures when a pair of technology classes co-occur more often than what would be expected under the assumption that this relation would be statistically independent. The indicator is normalized following the method of association strength, which according to Eck and Waltman (2009), is the best method among the most well-known ones for normalizing co-occurrence data (which is the case of relatedness).

---
[2] We use the EconGeo R-based package (Balland, 2017).

Moreover, we use the method of reflections for calculating complexity, as proposed in Hidalgo and Hausmann (2009). It considers the performance of the entity in a given output (e.g., the number of patents produced in a technological field by a country) and the ubiquity of the output considered (e.g., the number of countries exploring the same technology). We calculate complexity based on the implementation of the method of reflections proposed by Balland (2017) through the index knowledge complexity. We calculate this index for two distinct kinds of entities, namely technologies and countries, adapting the MORt function suggested for technologies (see Balland (2017), p. 42) and the MORc function for regions (see Balland (2017), p. 40), respectively. Moreover, the method produces a symmetric set of two types of variables (e.g., a country and its performance in each of the considered technologies) which can be further iterated according to a defined number of steps. Each interaction step (also called 'reflections') result in the average value of the previous-level properties of the considered variable. We apply one and two steps for the MORc and MORt calculations respectively, justifying each choice accordingly at the relevant stage of the subsequent empirical analysis.

We divide the following empirical analysis into two parts: the first is technological specific, where we analyse how AI evolves during the observation period (1974-2018); the second part is country-specific, where we evaluate how leading countries explore AI and the possible effects of this exploration in their technological space during this observation period. In both parts, we differentiate three 15-years intervals (1974 - 1988, 1989 - 2003, and 2014 - 2018) to evaluate changes during the full observation period.

In both parts of the analysis, we use the same 'global technological space', adapted from Hidalgo et al. (2007), which we create using information from all priority filings registered in the observation period (1974-2018). We use the 3$^{rd}$ version of the IPC technological field classification to differentiate the technologies in this global technological space. This classification facilitates international comparisons and overcomes some inconsistencies of earlier classifications. The selected technological field classification is useful for our purpose, as it considers all existent IPC codes, balances the size of the considered fields, and reduces the overlap between similar technologies (Schmoch, 2008).

Furthermore, we apply the Revealed Comparative Advantage (RCA) index to measure how AI and countries' related specialisations change over time. The RCA index, as presented in Balassa (1965), is widely used in knowledge space analyses based on patent data (Balland et al., 2019; Colombelli et al., 2013; Hidalgo et al., 2007; Petralia et al., 2017; Rigby, 2015). It relates the share of a given entity (e.g., technology, country) to the share of this same entity in a larger economy (e.g., overall technologies, countries). The basic premise is that if a given entity has a production level higher than the global average, this entity holds a comparative advantage in producing it. If the entity has an RCA equal or higher than one, it has a specialisation, whereas values below this threshold show an absence of specialisation. The complete implementation of our analysis is available in our GitHub repository[3].

## 4. Empirical analysis

### 4.1. Identifying the technological core of AI in the global technology space

In our dataset, 80.6% of AI's priority filings were registered between 2004 and 2018. This recent increase has been associated with the fact that AI expanded from specific applications in computer science to include a wide range of distinct functional applications involving diverse areas and industries including agriculture, healthcare, and manufacturing (Li & Jiang, 2017; WIPO, 2019). Even before this expansion, defining

---

[3] https://github.com/matheusleusin/Paper-The_Emergence_of_Artificial_Intelligence

what constitutes AI has been challenging (Lupu, 2018; Righi et al., 2020; Russell & Norvig, 2016).

Although precisely delimiting AI theoretically might be a challenging task, it seems feasible to us to determine empirically the 'technological core' of it. In this sense, we follow Dosi and Grazzi (2006) in arguing that technologies have a structure of technological knowledge over which patterns of technological innovation tend to be relatively invariant and linked through specific routes to the solution of particular problems. Thus, we aim at identifying the particular structure of technological knowledge used in AI applications, which is to be stable over time. We label this structure 'technological core', borrowing the term from the GPT theory, where it is used to describe a cluster of technologies that share some similarity and constitute the core of knowledge related to a GPT.

We start by creating a general structure of technological knowledge using all patents (i.e., not only AI) during the whole observation period (1974-2018). We create a 'global technological space', which we depict as a network, where the nodes represent technology fields and the edges between them their relatedness. By using a measure of similarity between these fields, we aim to identify the general structure of the technological knowledge related to these fields, as presented in Figure 1 (see Appendix B for a complete list of the fields' complexities)[4].

---

[4] Note that the label 'dgr' highlighted in Figure 1 stands for Degree of Complexity (i.e., higher values mean more complex technological fields), whereas the label 'weight' stands for the weight of the links (i.e., higher values mean technological fields connected to a higher number of other technological fields).

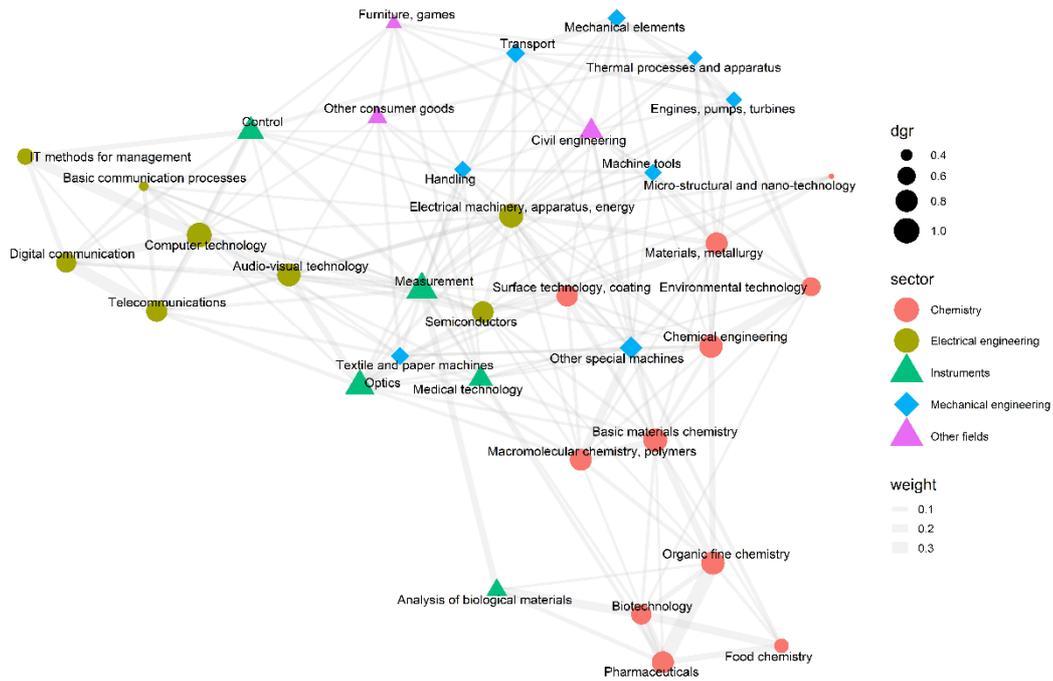

**Figure 1.** Global Technological Space for IPC fields.

As expected, the global technological space highlights differences between sector-related positions, meaning that technological fields within the same sector are often more related to each other than with other sectors. 'Electrical engineering'-related fields are mostly placed jointly on the left side of the network, while 'mechanical engineering' and 'chemistry'-related ones are placed on the top and bottom of the figure respectively (see Figure 1). The second most complex field of the network, 'electrical machinery, apparatus, energy' is placed in the centre, while the most complex field of 'computer technology' is placed on the left side.

We indicate over this global technological space which technological fields were most frequently adopted in AI innovations for each of the three 15-year periods (see Figure 2). We identify these fields by measuring the binary RCA index (i.e., the specialisation) of all AI innovations (i.e., all patents identified in our 'AI dataset') for each interval.

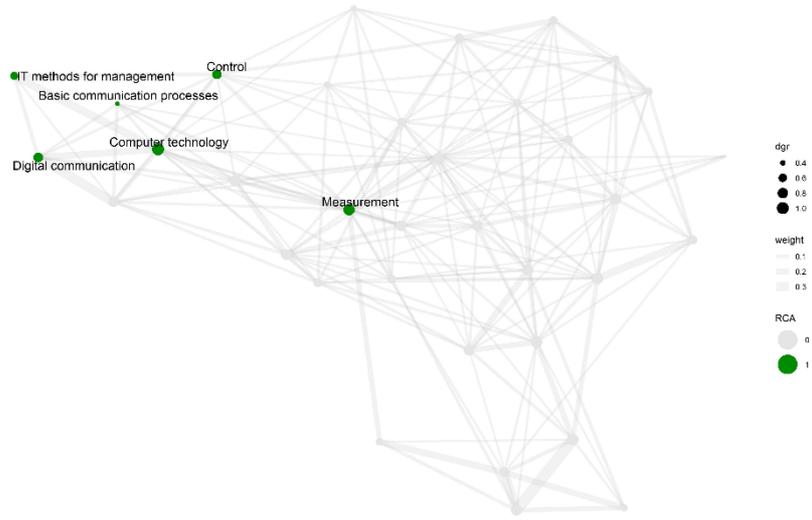
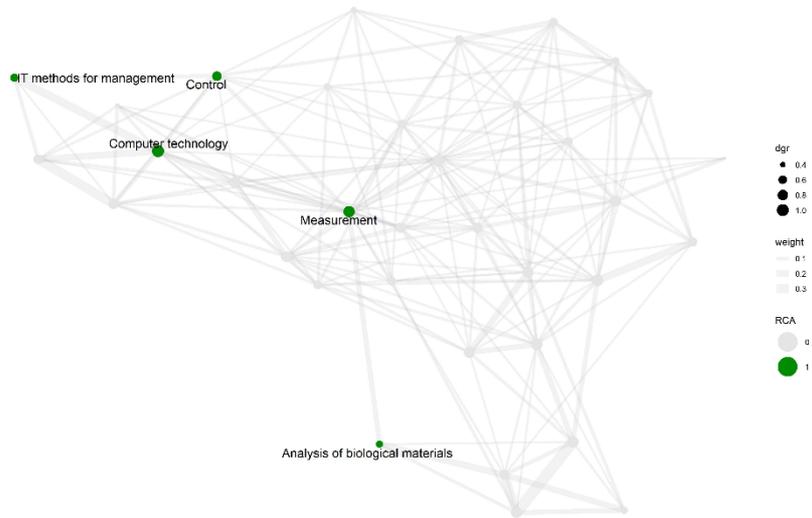
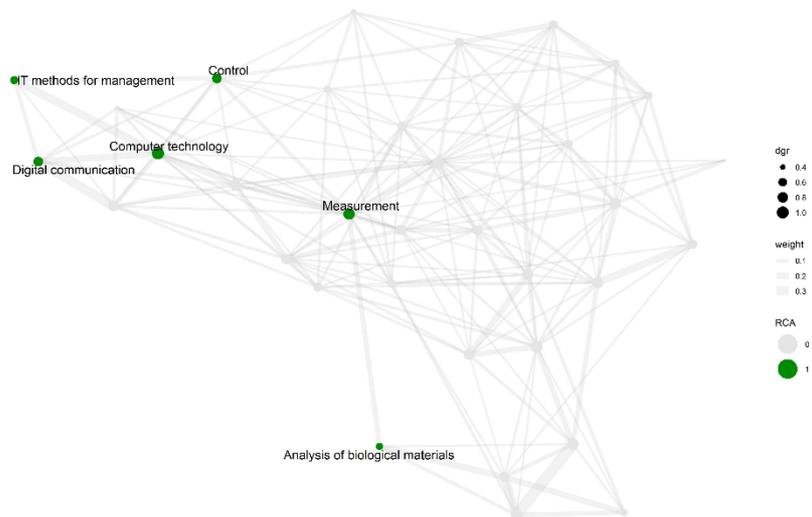

**Figure 2.** AI specialisations over the considered periods and IPC fields.

The technological fields most used in AI innovations are placed on the left side of the network, around the most complex 'computer technology' field (see Figure 2). All specialisations of AI are kept exclusively in the sectors of 'electrical engineering' and 'instruments'. Despite showing some specialisation in seven (20%) of the total of 35 technological fields available, AI's RCAs change slowly over time, maintaining a constant specialisation in the technological fields of 'computer technology', 'IT methods for management', 'control', and 'measurement', while moving from the left-side less complex 'basic communication processes' to include the right-sided and more complex field of 'analysis of biological materials'. We suggest that AI's technological core lies within four technological fields: Computer technology, IT methods for management, Control, and Measurement, due to their constant specialisation over the three considered intervals. We define technological fields as closely 'related to AI' when they present a specialisation in at least one interval. Furthermore, we define the category 'AI-surrounding fields' for technological fields that had no specialisation in AI in any period, but that are visually located around the AI cluster in the network. This implies that these technologies, in general, are more closely related to AI than others. We create this last category label by picking six technological fields that are close to AI-core fields and present a high degree of connections to these fields.

*4.2. Relatedness and knowledge complexity of AI in the global technology space*

Next, we measure how knowledge complexity and relatedness of AI-related inventions change over time overall and separately for the categories of 'AI-core fields', 'AI-related fields', and 'AI-surrounding fields'. For relatedness, we take the average relatedness for each of the previous AI categories selected by technological fields (see Appendix C). For complexity, we use iterations by adapting the MORt function for technological complexity (see Balland (2017)), which holds that zero iteration gives ubiquity, whereas one and two iterations give average diversity of regions that have an RCA in the technology and the average ubiquity of technologies developed in the same regions, respectively. As we assume AI as a cluster of related technologies connected (in a similar way to what occurs in regions), and since we are not interested in its diversity, we choose two steps for this calculation, summing the results over the related fields of each considered category (see Figure 3).

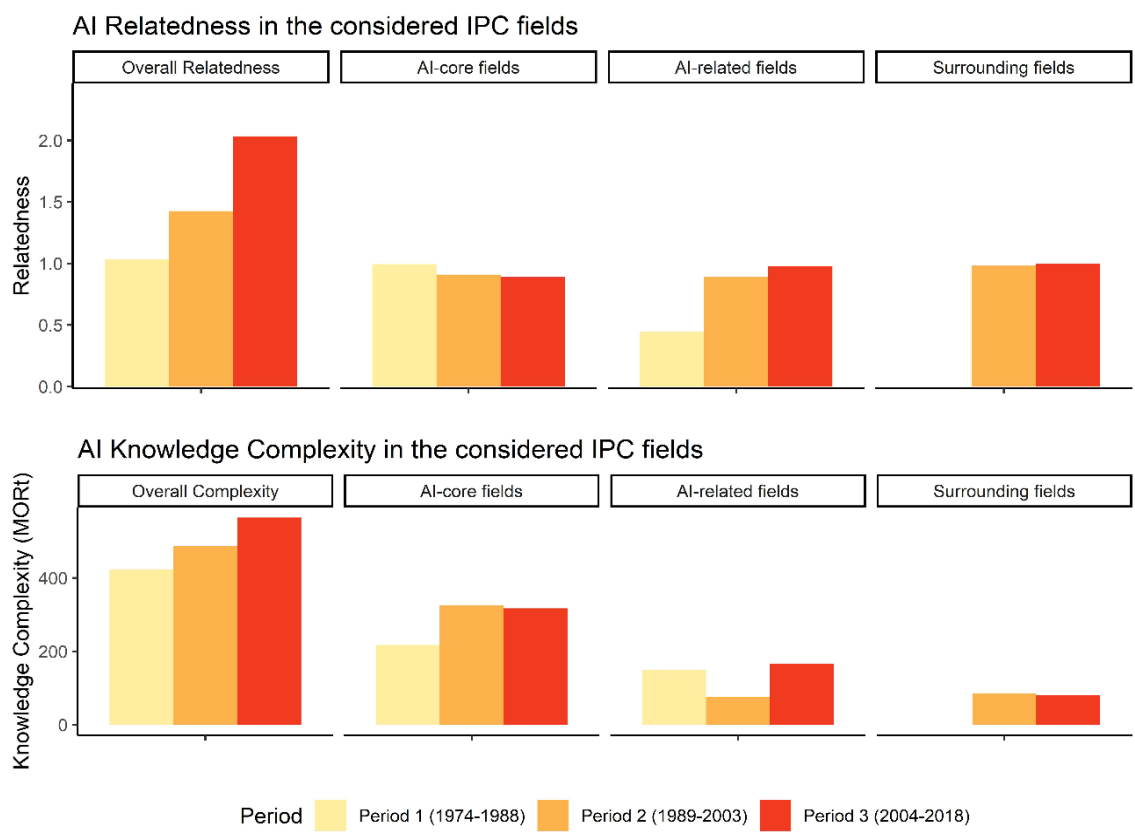

**Figure 3.** AI relatedness and knowledge complexity in the considered IPC fields.

We find from the global perspective that the overall relatedness and complexity of AI inventions increase over time, since the share of AI patents using the same technological fields increased in each subsequent period (increase in relatedness), as well as the average ubiquity of fields used in AI patents (increase in complexity). Yet, it is important to highlight that for AI-core fields relatedness actually decreases, while complexity increases. So, after the initial period, the fields composing AI's technological core are less used jointly in the same patents i.e. become more distinct from each other. In contrast, AI in related and surrounding technological fields follow the overall trend and increase in complexity and relatedness (with the exception of AI-related fields in the second interval). In sum, these results highlight two distinct trends for AI-related inventions: the majority of AI patents become more complex and related over time, while the technological core of AI – concentrated around the computer and measurement-related fields – become more complex but at the cost of decreasing its relatedness, meaning that at some point AI innovations based on AI-core technologies become less alike to each other.

*4.3. Specialisations in AI fields in the technology space at the country level*

More than 92% of the total number of AI priority filings in our dataset are accounted for by inventors from only four countries: China, Japan, South Korea, and the US (see Figure 4). Japan and the US are 'early leaders' in AI development, with some AI patents registered before 1988. South Korea and China emerge in the second period, with a particular spike in registers for China in the third period (see Figure 4). We analyse the technological development of these countries by outlining their overall technological specialisations over the three intervals in the global technological space, for all patents developed by these countries (see Figure 5). We further distinguish the respective categories of AI technological fields, as defined previously, in comparison to the remaining fields (labelled as 'Other').

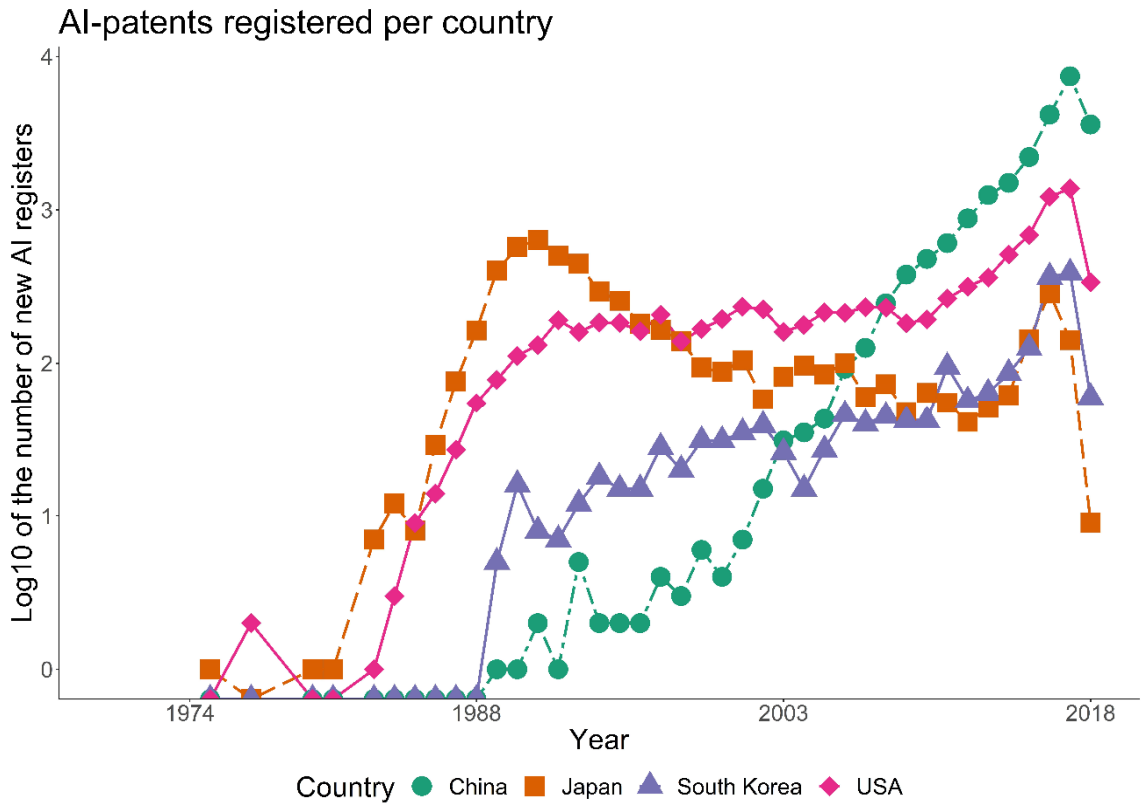

**Figure 4.** Log 10 of number of AI-priority filings by China, Japan, South Korea and the US.

For the technological core of AI, we record very distinct profiles of specialisations for the four countries (see Figure 5). The US, for example, starts with no specialisation in any AI-core fields, but gradually moves toward this technology's cluster, becoming specialised in most of the AI-core fields (the only exception being the field 'measurement'). Japan, on the other hand, starts with specialisation advantages in all of the AI-core fields and in some AI-related fields, but loses some of these specialisations each following period, while gradually moving towards the 'mechanical engineering' related part of the network. South Korea has most of its specialisations in the 'chemistry' sector in the first period, moves to the AI-related cluster in the second period, and loses the most complex 'computer technology' specialisation in the third period. Finally, China has most of its specialisations in the sector of 'chemistry' in the initial period, and moves very slowly to the left side of the network, dominating a few technological fields related to AI.

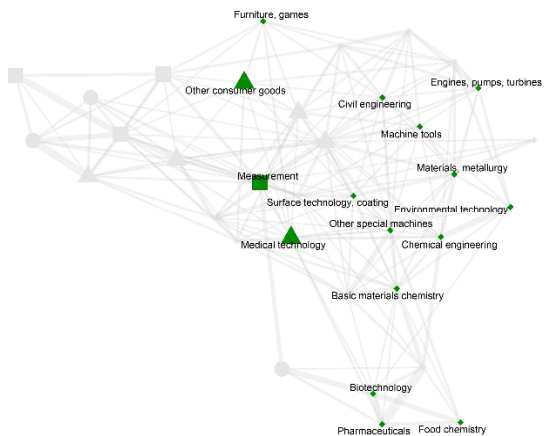 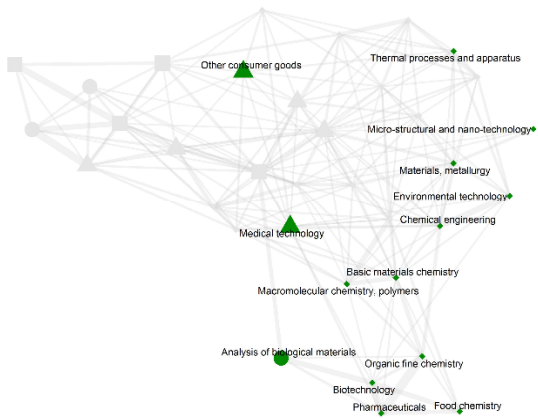 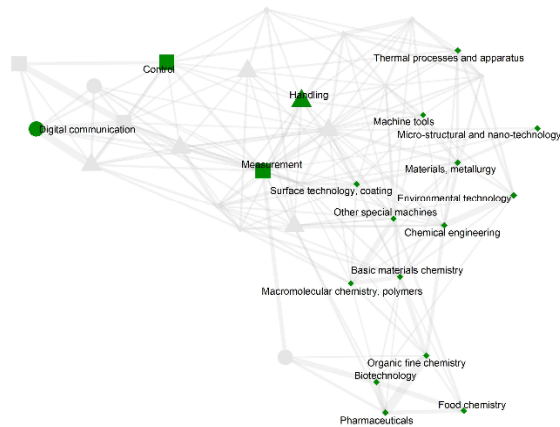 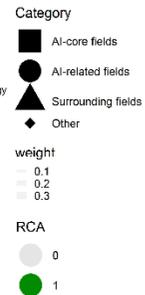

a)

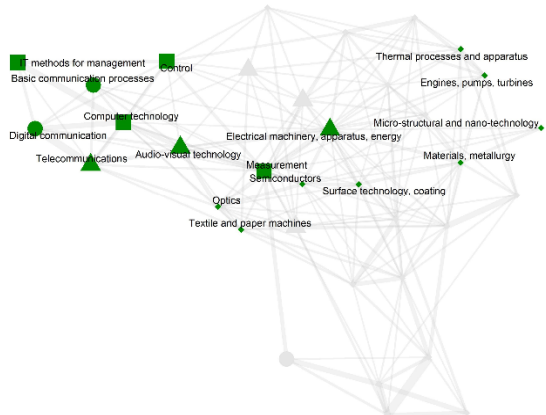 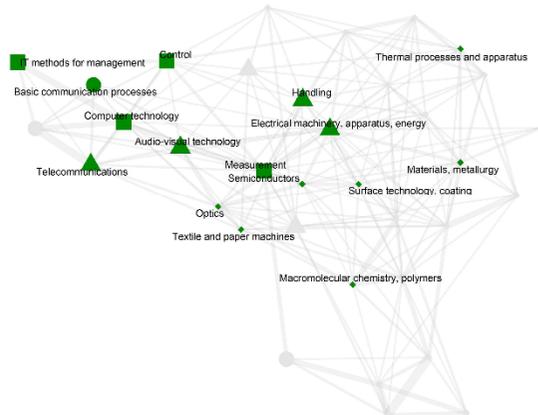 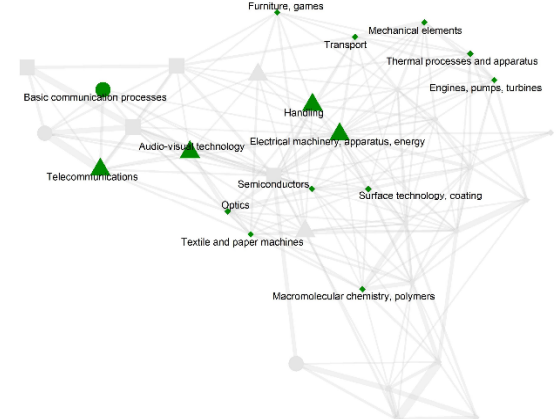 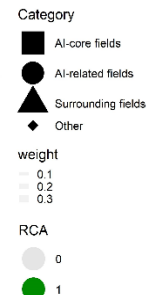

b)

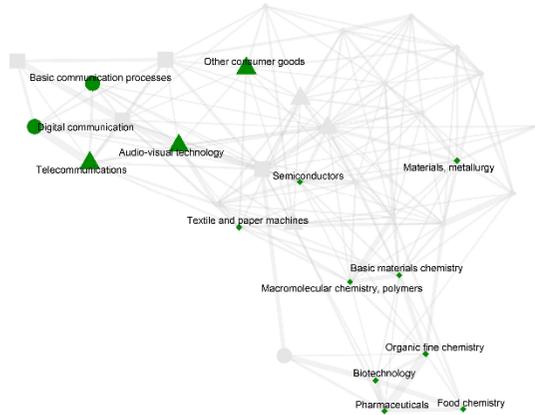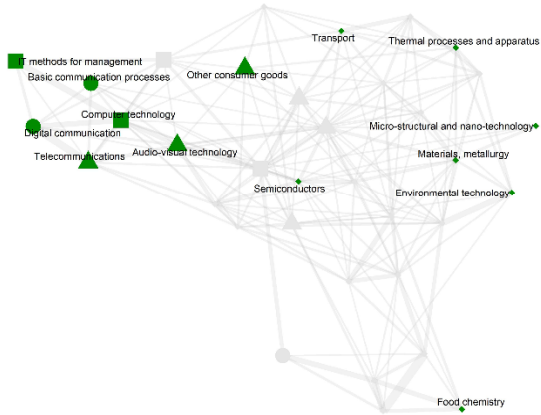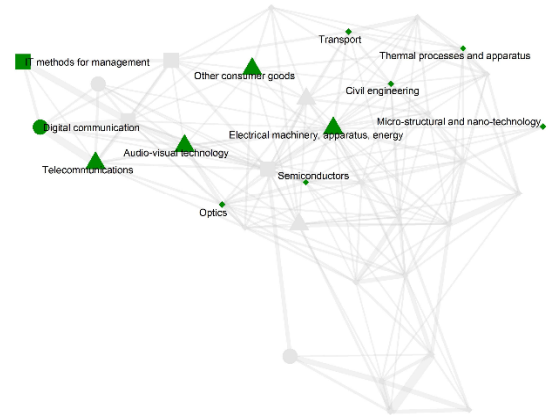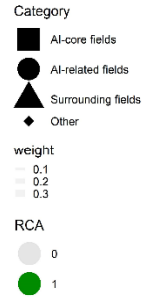

c)

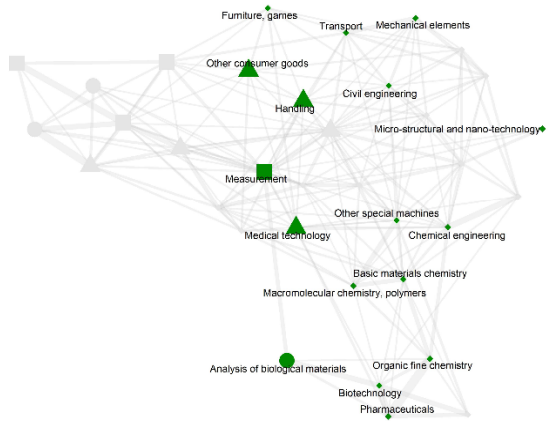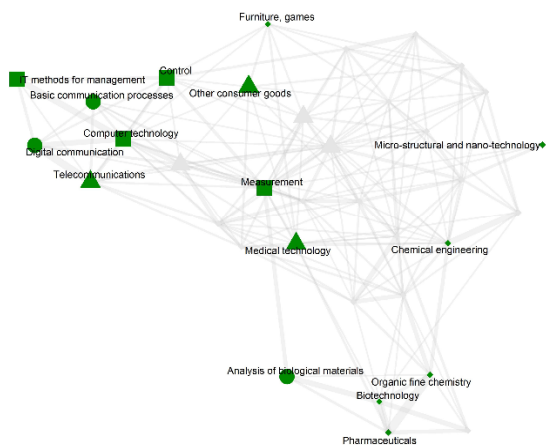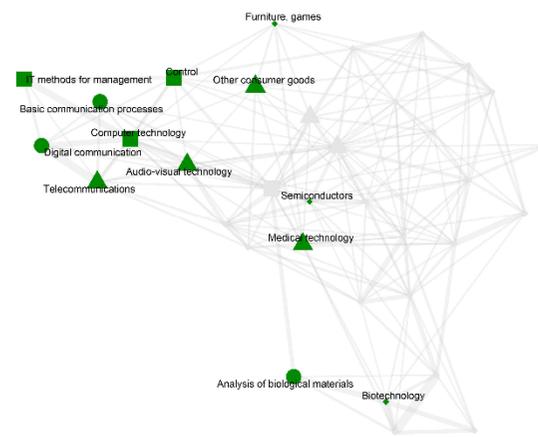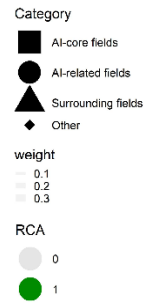

d)

**Figure 5.** Specialisations in the considered IPC fields for China (a), Japan (b), South Korea (c), and the United States (d).

*4.4. Overall and AI specialisation in technological classes at the country level*

Despite the increase in the number of technological fields used in AI technologies, most AI patents are still concentrated in a few IPC codes. In the following, we analyse countries' overall specialisation (see upper panel Figure 6) compared to their specialisation within AI (see lower panel Figure 6) across ten selected IPC subclasses. The ten selected IPC subclasses are the most frequently used in AI patents across the three intervals considered. They concentrate 77% of the subclasses used in AI registers; eight of them are within AI-core technological fields: subclasses G06F, G06K, G06N, G06T, G10L (technological field of 'computer technology'), subclass G01N ('measurement), subclass G05B ('control'), and subclass G06Q ('IT methods for management'); and the remaining two subclasses are within the AI-related field 'digital communication' (subclass H04L) and on the AI-surrounding field 'medical technology' (subclass A61B). The average RCA[5] values of the four countries are added for each code and period considered (see dotted lines in panels of Figure 6).

---

[5] This time we measure RCA as a continuous indicator and use the Log10 transformations of the corresponding values (i.e., there is a specialisation for values above 0).

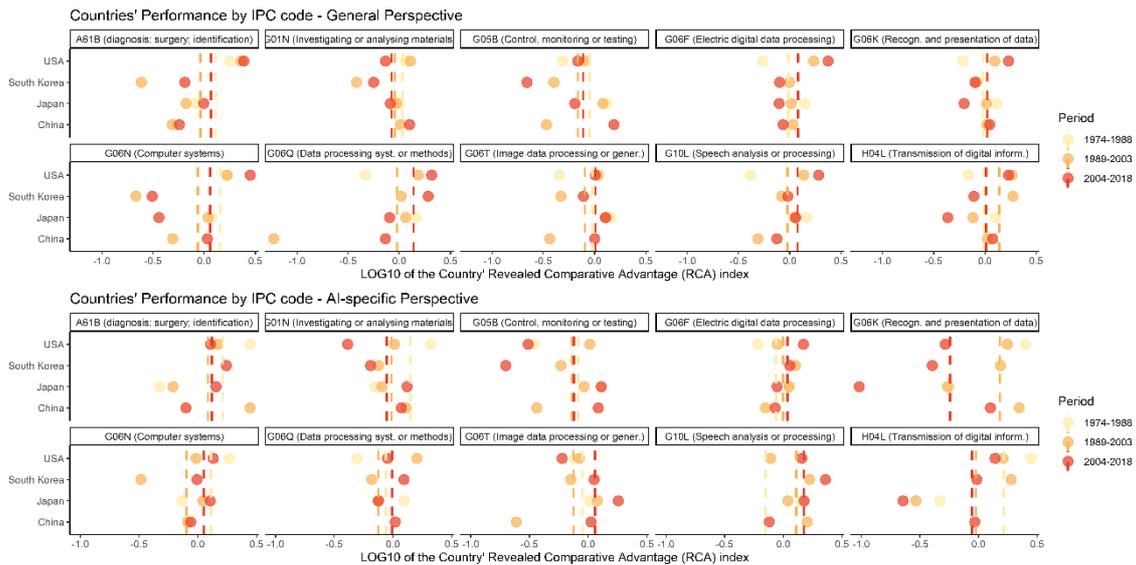

**Figure 6.** Overall and AI-specific RCA across AI patent subclasses for China, Japan, South Korea, and US.

In most cases, we find that the specialisation advantage in a subclass within a country's AI patents corresponds to a general specialisation advantage of the country in this subclass in the respective period (see Figure 6). This suggests that specialisations within AI are associated to countries' general specialisation profiles. We find rarely a specialisation within AI preceding a general specialisation in the same subclass. Where this is the case, it does not seem to last. For example, China loses very rapidly its new 'unrelated' A61B and G10L specialisations in the next period. The same holds for the US in the codes G05B, G06K, and H04L, as well as South Korea for G06K. The only exception to this is South Korea's sustained specialisation in the subclass 'Speech analysis or processing' (code G10L). We also detect that despite Japan losing several of its general specialisations in the third period in its national technological space (see Figure 5b), the country maintains specialisation advantages within AI in the same period (e.g., leading in the subclasses G05B and G06T as well as relevant in G10L, A61B, G01N, and G06N).

Furthermore, Japan and the US are the only countries with specialisation in the most complex 'computer technology' field in the third period in the national technological space perspective (see Figure 5b and d), and herein the 'computer systems' (code G06N) subclass (see Figure 6). With increasing specialisation in this area, both countries reduce their specialisations in the less complex 'recognition and presentation of data' subclass (code G06K), for example. The latecomers South Korea and China also move in the direction of the most complex 'G06N' subclass, although not reaching a specialisation in it yet. In sum, our observations indicate that leading countries not only develop AI according to their general existing knowledge capabilities, but also seem to move to more complex uses of AI as they further develop this technology.

*4.5. Technological relatedness and knowledge complexity at the country level*

Next, we measure changes in technological relatedness and complexity at the country level (see Figure 7). Thereby, we again differentiate between AI-core, AI-related, and AI-surrounding technological fields and apply iterations for calculating complexity in line with Balland (2017). As we are now interested in the average complexity of countries, i.e. the average ubiquity of their technological fields, we use one step for this calculation. For categories relatedness', we again take the average of each, while for complexity we don't apply any additional sum, since the MORc equation results in one value per country.

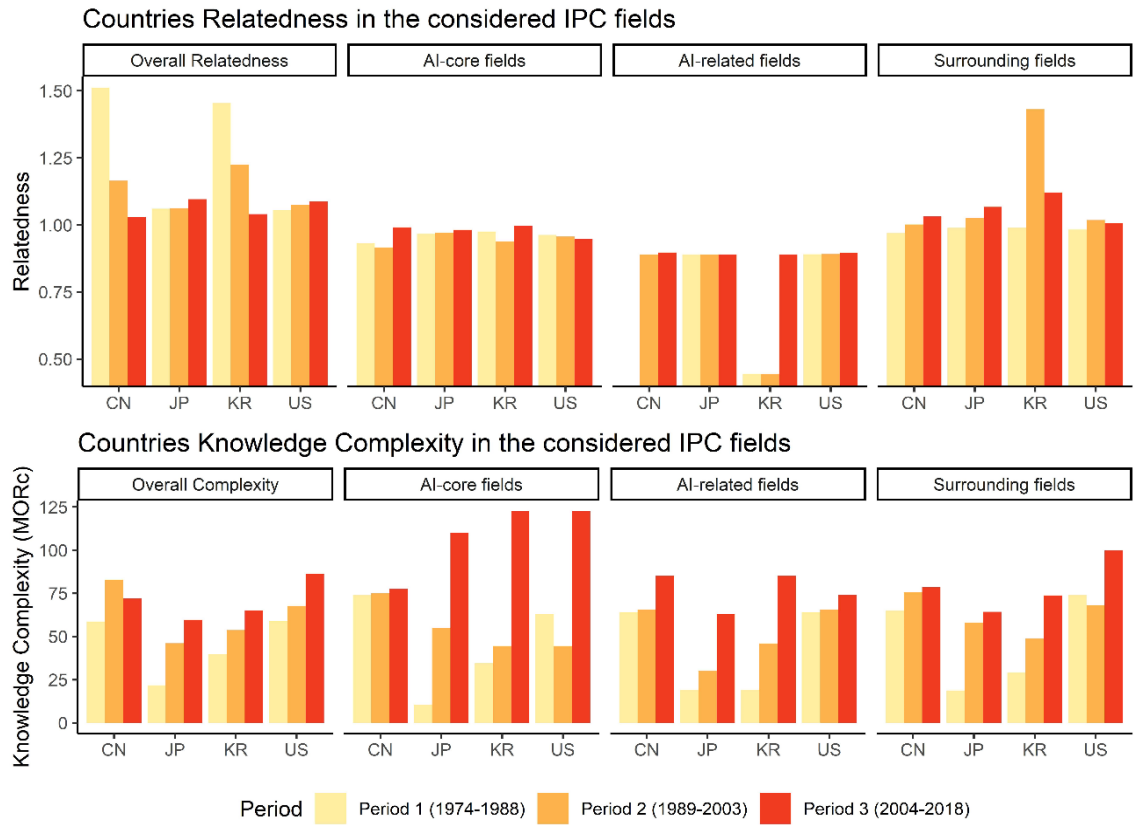

**Figure 7.** Relatedness and knowledge complexity in the considered IPC fields for China, Japan, South Korea and US.

We find distinct patterns of technological development at the country level. China, in a process of exploring and developing specialisations in a wide variety of fields, decreases the overall relatedness of its technological fields from interval to interval (see upper panel Figure 7). South Korea seems to follow the same trend. Conversely, the US and Japan increase their overall relatedness over the three intervals. Within AI categories, we find little difference in the trends for relatedness between countries. The only exception is South Korea in the AI surrounding fields in the second interval. This period coincides with the first time South Korea developed a specialisation in AI-core technologies ('computer technology' and 'IT methods for management') (see Figure 5c). This is the only period in which the country presents all of its AI specialisations in very closely related and connected technological fields.

For countries' knowledge complexity, there are indications for an association with the technological development of AI. In the case of Japan, which is an early AI leader and the only country having specialisations in every AI-core fields in the first interval, it happens also to be the country with the lowest overall knowledge complexity in this first period (see lower panel Figure 7). Over time, as AI development advances, the complexity of countries exploring it increases considerably. In fact, we find all countries have been increasing overall complexities over time, except for China in the second analysed period. When looking at the AI-core fields, China's rate of development is even less pronounced compared to the other three countries.

## 5. Discussion and concluding remarks

### 5.1. Summary of the main findings

At a global level, we delimitate AI's technological core from other technological fields, with the former being characterised as stable since the initial phase of AI development. We demonstrate that AI inventions have become both increasingly related and complex over time. When considering differences between the AI-core and the remaining categories, our results indicate that despite also increasing in complexity, AI inventions in its core technological fields have become more distinct. Figure 8 summarizes AI's development regarding the relatedness and complexity indicators for the main categories identified.

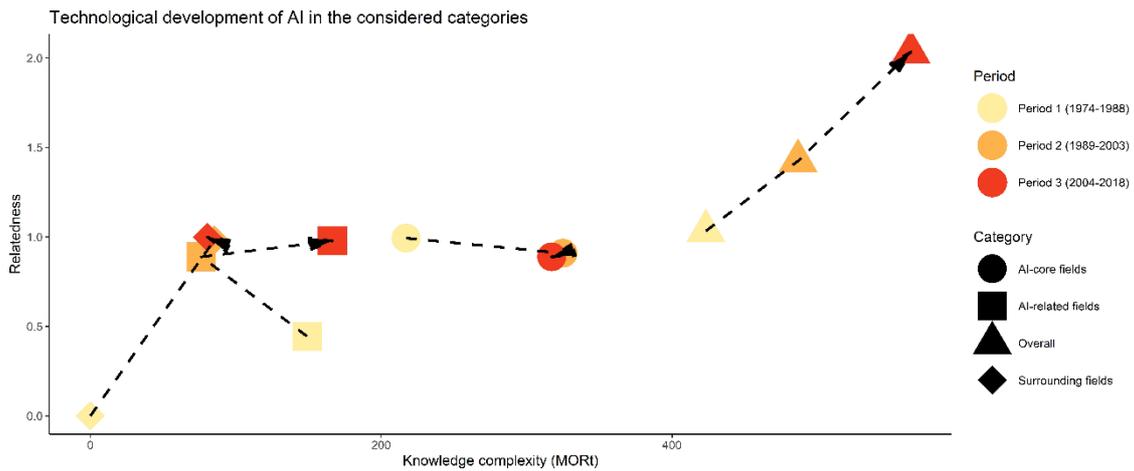

**Figure 8.** Relatedness and complexity of AI over time, for the considered categories. We find that AI overall has become a more complex technology since the first considered period, at least when comparing it to the remaining categories (see Figure 8). The larger increases in complexity stem from 'AI-core fields' and 'AI-related fields' in the second and third periods, respectively. The category 'surrounding fields' is the least complex one in almost all considered periods. Generally, we can summarize that AI technologies become overall more complex and related over time, confirming our related proposition.

At the country-level, our findings suggest technological relatedness as a potentially central factor in how countries develop AI. In addition to developing AI specialisations related to the already existing general specialisations, countries also move to more complex fields of AI as they further develop it. Moreover, it seems that AI-related specialisations developed in areas without previous general specialisation are rare and unlikely to persist. Conversely, AI-related specialisations developed by countries in a pre-existing general specialisation in most cases become persistent.

Figure 9 summarizes the overall technological development of the considered countries in terms of relatedness and complexity[6] indicators. Despite AI's increase in relatedness at the global level, countries developing AI do not present such a trend (see Figure 9).

---

6 Please note the distinction for the complexity indicator between a technological-related one for AI (MORt) and a geographical one for countries (MORc).

Japan and the US did slightly increase their overall relatedness. However, South Korea and China decreased it considerably.

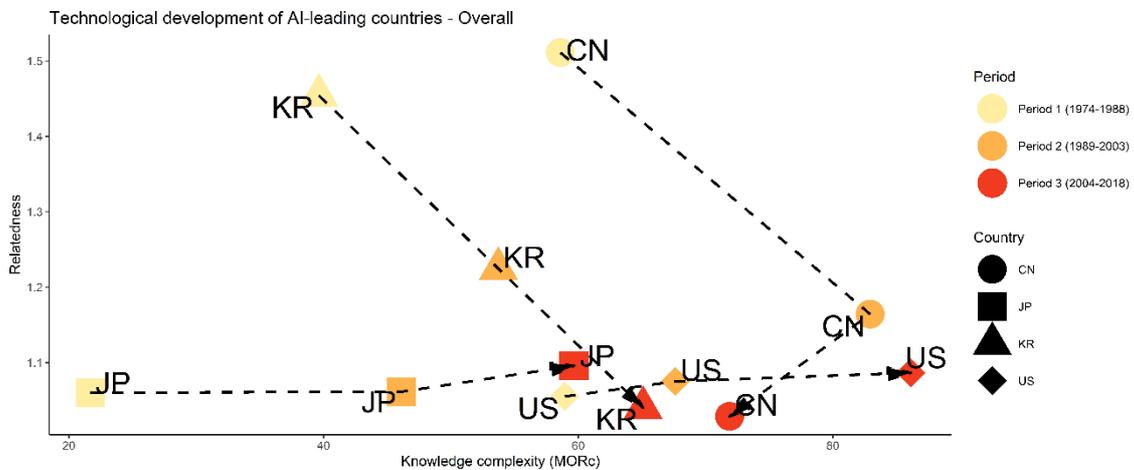

**Figure 9.** Overall relatedness and complexity of countries over time, for the considered leading countries.

For complexity, the association between AI emergence and countries' development is perhaps more apparent. Japan had most of its specialisations in technological fields related to AI in the first considered period and is the less technologically complex country in this period, when AI is still in an early development stage. As AI is further developed, the technological effects on countries' overall complexity become more explicit. In the third period, when AI reaches higher complexity, countries leading in overall complexity are the ones holding most specialisations in technological fields linked to the identified technological core of AI. The most complex US holds 3 of such specialisations, followed by China with 2, South Korea with 1, and Japan with none (as highlighted in Figure 5). Despite being last in the considered sample though, the early AI-leader Japan is also the country that most advanced in overall complexity in the considered periods, reaching a position close to the one held by South Korea and China in the last period. In sum, regarding our propositions related to countries, there is indicative evidence that the national development of AI indeed increased their overall complexity, whereas, we find no indicative support for countries' overall relatedness.

*5.2. Discussion of implications for the state-of-the-art*

5.2.1. Literature on the complexity and relatedness perspectives

Some of our results confirm prior findings on relatedness and complexity in the long-term development of technological paths: firstly, firms (Breschi et al., 2003) and countries (Balland, 2016; Hidalgo et al., 2007) develop new technological capabilities close to the already existing and related ones; secondly, unrelated specialisations are rare and prone to fail (Boschma et al., 2014; Castaldi et al., 2015); thirdly, more complex technologies tend to be of higher value (Balland et al., 2019; Hidalgo & Hausmann, 2009) and countries move to more complex technologies as they evolve their technological capabilities (Hidalgo & Hausmann, 2009; Petralia et al., 2017). However, despite the macro-level evidence that countries move to more complex technologies as they evolve technologically, the evidence at the micro-level indicates that firms develop new capabilities mainly through knowledge-relatedness (Breschi et al., 2003), without a direct consideration for complexity. This still leaves us with the question of how countries overall move to more complex technologies.

Reflections upon our findings would perhaps suggest the following explanation: firms create new technologies by using variations of the knowledge they have based on knowledge-relatedness. But as countries evolve, more complex and yet related knowledge will be available to be recombined, increasing the country's overall complexity. Hidalgo and Hausmann (2009) present an analogy of complexity by comparing countries' capabilities to a Lego game. In this analogy, each Lego brick represents a distinct capability and so the technological capabilities of a country are compared to a bucket of Lego bricks. As a child creates distinct Lego models with Lego bricks at hand, distinct technologies are developed in countries based on the capabilities available. More technologically complex countries have a bigger 'bucket' of capabilities and thus more possibilities of combination.

Our argument is that by having just a few capabilities available, less technologically developed countries can only produce a limited number of distinct technologies. Thus, they present less technological variety but higher overall relatedness. Possible Lego models vary as much as the few Lego bricks available. As countries evolve technologically, new capabilities are created. Even if the most complex capabilities stay out of reach initially, at some point they are added to the bucket because there are no other options available or because they become feasible through knowledge-relatedness. At the micro-level, firms are picking the bricks available that they know how to combine (i.e., based on knowledge-relatedness) and combining them over and over again to produce new models (i.e., technologies). If the country continues to develop technologically, more complex models will be created by firms due to the increase in the variety of bricks available or because the newly added bricks are both related and more complex.

This could explain why less technologically developed countries present higher overall relatedness than developed ones. As these countries evolve technologically, their overall relatedness is reduced because there is an increase in the range of technological possibilities available and, thus, firms produce technological combinations increasingly unrelated to the previous ones. This insight is complementary to Petralia et al. (2017), who propose that countries become less dependent on relatedness as they become more technologically developed. In our analysis, this is the case of South Korea and China, which are less developed technologically at the start of the observation period and thus present a higher level of overall relatedness. Technologically developed countries, such as Japan and the US in our analysis, have a wide variety of capabilities available and, thus, also present lower overall relatedness.

The consideration of these technological dynamics might also provide a possible explanation as to why our preposition for relatedness could not be supported at the country level of analysis: we initially argue that countries extensively developing AI would increase their overall relatedness, since AI is a transversal technology in nature (Righi et al., 2020). So, innovations of a country developing AI extensively would become increasingly similar, due to knowledge-relatedness, as the AI technology is repeatedly combined with others across distinct technological fields. However, our analysis indicates that this was only the case for Japan and the US, whereas in the case of the less technologically developed countries South Korea and China, the overall relatedness was actually reduced. Thus, the proposed increase in the overall relatedness is possibly contingent upon the country's level of technological development. Only a sufficient level of technological development might be associated with a technological path towards a complex technology, which simultaneously favours increasing a country's overall relatedness.

### 5.2.2. Literature on AI

In terms of empirical insights into the long-term development of AI, our investigation contributes to the identification of technological fields related to AI. In contrast to earlier studies (Fujii & Managi, 2018; Klinger et al., 2018; Righi et al., 2020), we manage to link the development of AI to a stable set of 'core technologies'. We demonstrate that the technological core of AI refers to technological fields such as 'IT methods for management', 'control', and 'measurement', besides the expected 'computer technology', since its early development in the '70s. This highlights the early use of IT in AI innovations, rather than ICT as pointed out as the enabler of AI during the '90s (Righi et al., 2020). In this way, the diffusion of ICT technologies was an enhancer of AI development, similar to the development of high-performance parallel computing chips and large datasets which are recognized as such in the AI literature (Klinger et al., 2018; Li & Jiang, 2017).

The stability of this diversified 'technological core' also highlights the importance of considering the transversal nature of AI (Righi et al., 2020) in corresponding identification strategies. Earlier approaches (see for example Fujii and Managi (2018)), identify AI by collecting patent registers related to the 'G06N' patent class, which refers to 'computer systems based on specific computational models'. Our approach suggests that this identification might be too restrictive and potentially misleading, since countries can specialise in these patent classes without developing an AI specialisation (as seen for example, for Japan and China in our analysis). In this respect, the proposed keyword-based searches may have an advantage over identification strategies based on existing classifications.

Furthermore, prior research highlights distinct patterns of AI development in leading countries such as the US, China, South Korea, and Japan (Fujii & Managi, 2018; Klinger et al., 2018; Leusin et al., 2020; Righi et al., 2020). Our investigation adds to that by showing that countries leading in terms of AI inventions also increase the complexity of these AI inventions. In this sense, our study indicates that countries developing AI start with more simple inventions and move potentially through a similar technological path in the direction of more complex uses. Thus, to understand the long-term emergence of AI empirically, it is important to consider not only trends in AI outputs but also quality aspects of these outputs such as their complexity.

Moreover, Klinger et al. (2018) highlight the possible closing of the window of opportunity for new entrants in AI development. The relatedness perspective adopted in this paper allows deepening this discussion. Our data show that, after the pioneering exploration of AI by Japan and the US in the '70s and '80s, both South Korea and China joined the AI development in the late '80s. These entries coincided with the second 'AI winter' (1987-1993), in which the limitations of the then-popular knowledge-based expert systems became evident. In the following years, machine learning models were further developed, with a decisive advancement reached with the proposal of deep learning in Hinton and Salakhutdinov (2006), which ultimately marked the decline of the logic-based models and the evolution of AI towards data-driven applications (WIPO, 2019). It seems that South Korea and China have taken advantage of a unique window of opportunity caused by this change of paradigm.

Yet the relatedness perspective allows disentangling whether such a window favoured particularly these latecomers. Although South Korea managed to benefit from its existing technological capabilities to explore AI, China had very few capabilities in domains directly related to AI, before and after this window opened. Hence, China did not have a particularly favourable technological environment to explore AI when the window opened, and nevertheless managed to do it through AI-peripheral and simpler technological fields closer to the country's existing specialisations. This points to several 'entry points' possible for latecomer countries to engage in AI development via their current specialisations, which might be a benefit of the transversal nature of AI (Righi et al., 2020).

Since our indicative findings suggest that the window of opportunity to explore AI will remain open, a smart specialisation policy (Foray et al., 2011) might be a possible option for latecomers by identifying opportunities that countries can explore by extending existing specialisations to more complex and yet related technologies (see also Balland et al. (2019)). Such a strategy might allow less technologically developed countries to effectively deploy AI nationally, even if this technology continues to increase its complexity, once not all technological fields related to AI increase their complexity at the same rate.

*5.3. Limitations and future research directions*

We need to acknowledge the limitations of our approach, of which the most explicit ones are linked to the data. Firstly, we identify AI innovations by just considering patents, although many innovations in this field are not patented. This might constitute a particular issue for AI innovations based on software and even more to open source software development. Hence, in a strict sense, our results refer to AI 'inventions' (with proprietary characteristics) rather than 'innovations'. Secondly, we consider both granted and non-granted patents, which possibly introduce a quality bias on our dataset. Thirdly, we develop a keyword-based search to identify AI, which we prefer in our case over strategies based on classification codes. However, such an approach is inherently subjective and sensitive to the choice of keywords.

It is also important to highlight that the adopted methods for analyses presented in this study do not allow conclusions regarding causality. In particular, this applies to the 'effects' of countries' general specialisation advantages in AI specialisations as well as 'effects' or 'impacts' of AI's evolution on the technological paths of countries. These causality aspects are subject to further inspection, which would need to apply causal inference strategies to test, whether the indicative findings revealed in this study are robust and beyond reasonable econometric doubt. This might also require a larger set of countries under investigation during the observation period. Our study focused on four countries, which account for the lion's share of AI patents during the observation. Obviously, this choice limits our ability to generalize our findings.

Further, the use of the IPC technological field classification, which comprehends the 35 technological fields considered in our study, might hide relevant developments at the more granular level. For example, Petralia et al. (2017) present an analysis at the IPC class level (129 distinct codes) and find evidence that countries move through clear technological paths. Our study, however, points to less clear dynamics, with countries changing their technological specialisations considerably in each period. In fact, a possible explanation for this might be the difference in the aggregation in the employed technological classification. Also, a consideration of the individual characteristics of the distinct technological fields that compose the core of AI might be relevant for advancing the research on the catching up of latecomers. Here, prior studies point out that characteristics such as technological cycle time and differences between stocks of knowledge of distinct technological classes affect the catching up of latecomers (Park &

Lee, 2006). Finally, we also do not consider collaborations in the development of innovations, despite the evidence that it is an important factor for increasing complexity (van der Wouden, 2020).

**Appendix A. Search strategy applied in PATSTAT 2019 spring version for identifying AI patents.**

**Select** appln_id from tls202_appln_title
**Where** *appln_title* like '%machine learn%' OR *appln_title* like '%Probabilistic reason%' OR *appln_title* like '%Fuzzy logic%' OR *appln_title* like '%Logic Programming%' OR *appln_title* like '%Ontology engineer%' OR *appln_title* like '%pervised learn%' OR *appln_title* like '%reinforced learn%' OR *appln_title* like '%task learn%' OR *appln_title* like '%neural network%' OR *appln_title* like '%deep learn%' OR *appln_title* like '%expert system%' OR *appln_title* like '%support vector machin%' OR *appln_title* like '%description logistic%' OR *appln_title* like '%classification tree%' OR *appln_title* like '%regression tree%' OR *appln_title* like '%logical learn%' OR *appln_title* like '%relational learn%' OR *appln_title* like '%probabilistic graphical model%' OR *appln_title* like '%rule learn%' OR *appln_title* like '%instance-based learn%' OR *appln_title* like '%latent represent%' OR *appln_title* like '%bio-inspired approach%' OR *appln_title* like '%probability logic%' OR *appln_title* like '%probabilistic logic%' OR *appln_title* like '%reinforcement learn%' OR *appln_title* like '%multitask learn%' OR *appln_title* like '%Decision tree learn%' OR *appln_title* like '%support vector network%' OR *appln_title* like '%deep structured learn%' OR *appln_title* like '%hierarchical learn%' OR *appln_title* like '%graphical model%' OR *appln_title* like '%structured probabilistic model%' OR *appln_title* like '%Rule induction%' OR *appln_title* like '%memory-based learn%' OR *appln_title* like '%bio-inspired comput%' OR *appln_title* like '%biologically inspired comput%'

**UNION**
**Select** appln_id from tls203_appln_abstr
**Where** *appln_abstract* like '%machine learn%' OR *appln_abstract* like '%Probabilistic reason%' OR *appln_abstract* like '%Fuzzy logic%' OR *appln_abstract* like '%Logic Programming%' OR *appln_abstract* like '%Ontology engineer%' OR *appln_abstract* like '%pervised learn%' OR *appln_abstract* like '%reinforced learn%' OR *appln_abstract* like '%task learn%' OR *appln_abstract* like '%neural network%' OR *appln_abstract* like '%deep learn%' OR *appln_abstract* like '%expert system%' OR *appln_abstract* like '%support vector machin%' OR *appln_abstract* like '%description logistic%' OR *appln_abstract* like '%classification tree%' OR *appln_abstract* like '%regression tree%' OR *appln_abstract* like '%logical learn%' OR *appln_abstract* like '%relational learn%' OR *appln_abstract* like '%probabilistic graphical model%' OR *appln_abstract* like '%rule learn%' OR *appln_abstract* like '%instance-based learn%' OR *appln_abstract* like '%latent represent%' OR *appln_abstract* like '%bio-inspired approach%' OR *appln_abstract* like '%probability logic%' OR *appln_abstract* like '%probabilistic logic%' OR *appln_abstract* like '%reinforcement learn%' OR *appln_abstract* like '%multitask learn%' OR *appln_abstract* like '%Decision tree learn%' OR *appln_abstract* like '%support vector network%' OR *appln_abstract* like '%deep structured learn%' OR *appln_abstract* like '%hierarchical learn%' OR *appln_abstract* like '%graphical model%' OR *appln_abstract* like '%structured probabilistic model%' OR *appln_abstract* like '%Rule induction%' OR *appln_abstract* like '%memory-based learn%' OR *appln_abstract* like '%bio-inspired comput%' OR *appln_abstract* like '%biologically inspired comput%'

**Appendix B. Technological fields complexity considered in the global technological space network.**

| Field name | Sector | Degree |
|---|---|---|
| **Computer technology** | Electrical engineering | 1,000 |
| **Electrical machinery, apparatus, energy** | Electrical engineering | 0,955 |
| **Basic materials chemistry** | Chemistry | 0,938 |
| **Organic fine chemistry** | Chemistry | 0,913 |
| **Measurement** | Instruments | 0,912 |
| **Chemical engineering** | Chemistry | 0,906 |
| **Audio-visual technology** | Electrical engineering | 0,901 |
| **Other special machines** | Mechanical engineering | 0,856 |
| **Optics** | Instruments | 0,839 |
| **Pharmaceuticals** | Chemistry | 0,825 |
| **Materials, metallurgy** | Chemistry | 0,822 |
| **Macromolecular chemistry, polymers** | Chemistry | 0,820 |
| **Semiconductors** | Electrical engineering | 0,804 |
| **Surface technology, coating** | Chemistry | 0,790 |
| **Telecommunications** | Electrical engineering | 0,782 |
| **Biotechnology** | Chemistry | 0,738 |
| **Digital communication** | Electrical engineering | 0,719 |
| **Control** | Instruments | 0,708 |
| **Transport** | Mechanical engineering | 0,688 |
| **Environmental technology** | Chemistry | 0,670 |
| **Textile and paper machines** | Mechanical engineering | 0,631 |
| **Mechanical elements** | Mechanical engineering | 0,615 |
| **Medical technology** | Instruments | 0,611 |
| **Civil engineering** | Other fields | 0,607 |
| **Machine tools** | Mechanical engineering | 0,602 |
| **Handling** | Mechanical engineering | 0,601 |
| **IT methods for management** | Electrical engineering | 0,566 |
| **Engines, pumps, turbines** | Mechanical engineering | 0,545 |
| **Analysis of biological materials** | Instruments | 0,510 |
| **Thermal processes and apparatus** | Mechanical engineering | 0,508 |
| **Other consumer goods** | Other fields | 0,499 |
| **Food chemistry** | Chemistry | 0,491 |
| **Furniture, games** | Other fields | 0,427 |
| **Basic communication processes** | Electrical engineering | 0,376 |
| **Micro-structural and nano-technology** | Chemistry | 0,346 |

**Appendix C. IPC fields considered for each category.**

| Description | Category considered |
|---|---|
| **Computer technology** | AI-core fields |
| **IT methods for management** | AI-core fields |
| **Measurement** | AI-core fields |
| **Control** | AI-core fields |
| **Analysis of biological materials** | AI-related fields |
| **Basic communication processes** | AI-related fields |
| **Digital communication** | AI-related fields |
| **Telecommunications** | Surrounding fields |
| **Audio-visual technology** | Surrounding fields |
| **Electrical machinery, apparatus, energy** | Surrounding fields |
| **Medical technology** | Surrounding fields |
| **Handling** | Surrounding fields |
| **Other consumer goods** | Surrounding fields |